# Physical Uplink Control Channel Design for 5G New Radio

Lopamudra Kundu, Gang Xiong, Joonyoung Cho
Next Generation and Standards
Intel Corporation
{Santa Clara, Hillsboro, Hillsboro}, USA
Email: {lopamudra.kundu, gang.xiong, joonyoung.cho}@intel.com

*Abstract*—The next generation wireless communication system, 5G, or New Radio (NR) will provide access to information and sharing of data anywhere, anytime by various users and applications with diverse multi-dimensional requirements. Physical Uplink Control Channel (PUCCH), which is mainly utilized to convey Uplink Control Information (UCI), is a fundamental building component to enable NR system. Compared to Long Term Evolution (LTE), more flexible PUCCH structure is specified in NR, aiming to support diverse applications and use cases. This paper describes the design principles of various NR PUCCH formats and the underlying physical structures. Further, extensive simulation results are presented to explain the considerations behind the NR PUCCH design.

*Index Terms*— 5G, NR, PUCCH

## I. INTRODUCTION

Mobile communication has evolved significantly from early voice systems to today's highly sophisticated integrated communication platform. The next generation wireless communication system, 5G, or New Radio (NR), which is currently being developed by the standard organization, Third Generation Partnership Project (3GPP), will provide access to information and sharing of data anywhere, anytime by various users and applications [1][2]. As a unified network and system, NR is targeted to meet diverse multi-dimensional requirements, which are driven by different services and applications, including enhanced Mobile Broadband (eMBB), massive Machine Type Communications (mMTC) and Ultra-Reliable and Low Latency Communications (URLLC) [3].

As a fundamental building component to enable NR system, physical Uplink Control Channel (PUCCH) is mainly utilized to convey Uplink Control Information (UCI), including [4]:

- HARQ-ACK (Hybrid Automated Repeat Request-Acknowledgement) feedback in response to downlink data transmission.
- Scheduling Request (SR) which is used to request resource for uplink data transmission.
- Channel State Information (CSI) report which is used for link adaptation and downlink data scheduling. More specifically, CSI report may include Channel Quality Indicator (CQI), Pre-coding Matrix Indicator (PMI), Rank Indicator (RI), Layer Indicator (LI) and beam related information.

In LTE, PUCCH is transmitted in one or more Physical Resource Blocks (PRB) at the edges of the system bandwidth, following a mirrored pattern with slot level frequency hopping within a subframe so as to maximize the frequency diversity [5]. In NR, more flexible PUCCH structures need to be considered towards targeting different applications and use cases, especially for the support of low latency application such as URLLC.

Fig. 1 illustrates NR PUCCH structure with short and long durations. More specifically, NR PUCCH with short duration spans 1 or 2 symbol(s) in a slot, and can be multiplexed with Downlink (DL) and Uplink (UL) data channels in a Time Division Multiplexing (TDM) manner. Further, PUCCH can be inserted in the last part of one slot to enable fast HARQ-ACK feedback.

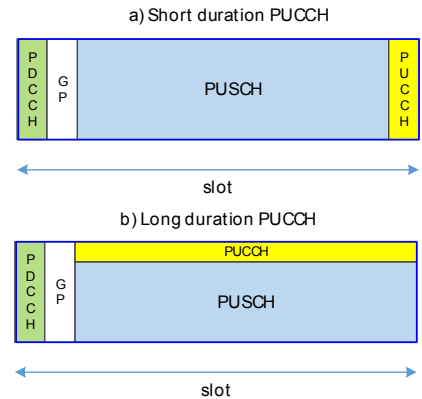

Fig. 1. Short and long duration NR PUCCH

Given that PUCCH carries critical information (e.g., HARQ-ACK feedback), a robust NR PUCCH design is necessary. Towards this end, multiple OFDM symbols can be allocated for NR PUCCH with long duration so as to ensure adequate coverage and robustness. As shown in the Fig. 1, NR PUCCH and Physical Uplink Shared Channel (PUSCH) can be multiplexed in a Frequency Division Multiplexing (FDM) fashion. Note that, for Time Division Duplex (TDD) system, depending on DL control resource set size, Guard Period (GP) duration and whether short PUCCH or Sounding Reference Signal (SRS) is present in the same slot, the starting symbol and duration for long PUCCH transmission can vary substantially. Hence, the design for long duration PUCCH should be scalable in time domain. In NR, the number of symbols for long PUCCH in a slot can range from 4 to 14.

As both short and long duration PUCCHs are supported, multiple PUCCH formats are specified in NR to carry various UCI payload sizes on different formats appropriate for the deployment scenarios and use cases, as listed in TABLE I.

TABLE I. NR PUCCH FORMATS

| PUCCH format | Short / long duration | Length (symbols) | Number of UCI bits | Number of PRBs |
|---|---|---|---|---|
| 0 | Short | 1~2 | 1, 2 | 1 |
| 1 | Long | 4~14 | 1, 2 | 1 |
| 2 | Short | 1~2 | > 2 | 1~16 |
| 3 | Long | 4~14 | > 2 | 1~6, 8~10, 12, 15, 16 |
| 4 | Long | 4~14 | > 2 | 1 |

The remainder of the paper is organized as follows. Section II describes the short duration PUCCH structure including PUCCH format 0 and format 2. In Section III, we present the design of long duration PUCCH for PUCCH formats 1, 3 and 4. Finally, we conclude the paper in Section IV.

## II. SHORT DURATION PUCCH

Short duration PUCCH can span 1~2 symbols within a slot and may carry either small (1~2 UCI bits) or large (more than 2 UCI bits) payload size. Short PUCCH carrying small payload size is referred to *PUCCH format 0* (PF 0), while the one carrying large payload size is called *PUCCH format 2* (PF2).

The detailed structures and design approaches of short duration NR PUCCH formats, i.e. PF 0 and PF 2 are described in the following subsections.

### A. PUCCH Format 0

NR PF0 spans over 1~2 symbol(s) in a slot and carries typically 1~2 UCI bits, which may be either HARQ-ACK bit(s) or SR bit or both. Two different approaches can be considered for the design of PF 0, e.g. a) Demodulation Reference Signal (DMRS) based structure and b) sequence based structure, as illustrated in Fig. 2.

*a) DMRS Based Structure:* For this design approach, DMRS is embedded in UL control channel during UCI transmission so that gNB (NR NodeB) can coherently demodulate HARQ-ACK at the receiver [6]. For DMRS insertion, FDM based multiplexing of DMRS and UCI symbols is used. Computer generated, low PAPR (Peak-to-Average Power Ratio) sequence is used for both DMRS and spreading sequence for Binary Phase Shift Keying (BPSK) (1-bit HARQ-ACK) or Quadrature Phase Shift Keying (QPSK) (2-bit HARQ-ACK) modulated UCI symbols. One example of DMRS based structure is shown in Fig. 2a where DMRS and UCI symbols are interleaved over 2 PRBs in frequency domain, each having 12 Resource Elements (REs) and DMRS overhead (i.e. the ratio of number of DMRS symbols to the total number of DMRS and UCI symbols) of 1/2.

*b) Sequence Based Structure:* In lieu of DMRS based structure, sequence based structure can be employed, where DMRS overhead can be eliminated completely since channel estimation is not required for non-coherent detection used in sequence based structure [6]. In this scheme, independent resources in the code domain can be assigned for HARQ-ACK feedback, as shown in Fig. 2b.

For 1-bit information, two independent resources are allocated for information bits '1' and '0' respectively, whereas for 2-bit information, four independent resources are required for transmission of information bits 00, 01, 11 and 10, respectively, using gray code. Independent resources correspond to different cyclic shifted versions of a computer generated, low PAPR sequence, which are orthogonal in frequency (i.e. zero cross-correlation). In this sequence based structure, gNB performs simple energy detection to differentiate Acknowledgement/Negative Acknowledgement (ACK/NACK) and thus leads to reduced receiver complexity. Unlike DMRS based structure, the sequence based structure is not affected by channel estimation impairment.

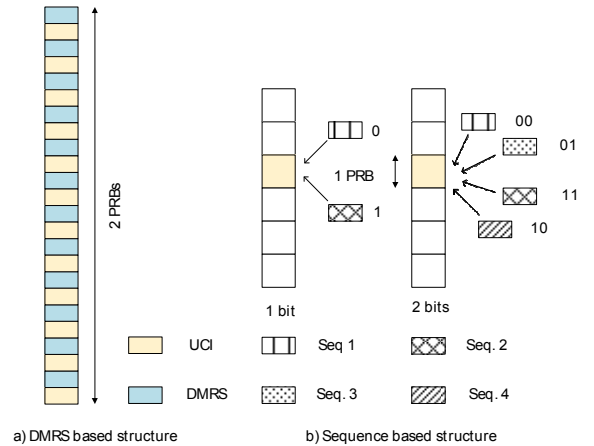

Fig. 2. Candidate structures for PUCCH format 0

*c) Simulation of DMRS and Sequence Based Structures:* Mathematically, DMRS based structure can be modeled as BPSK (1-bit) or QPSK (2-bit) modulation as explained above, while the sequence based structure follows the structural property of non-coherent orthogonal M-ary Frequency Shift Keying (FSK) modulation with M = 2 (for 1 bit) and M = 4 (for 2 bits) [6]. Therefore, the Bit Error Rate (BER) curves of DMRS and sequence based structures plotted as a function of normalized Signal-to-Noise Ratio (SNR) or SNR-per-bit ($E_b/N_0$) follow the trend of these traditional modulation schemes, as depicted in Fig. 3 for Additive White Gaussian Noise (AWGN) channel with ideal channel estimation.

For fading channels, however, the performance of both DMRS and sequence based structures degrade, and for highly frequency-selective channels (under large delay spread, for example), the degradation is more severe for DMRS based structure due to channel estimation error. In fact, for the case of 2-bit HARQ-ACK, the sequence based structure may even outperform DMRS based structure, as shown in Fig. 4.

In this link level simulation, DTX (Discontinuous Transmission) threshold (i.e. DTX-to-ACK error rate) is maintained at around $10^{-2}$ for both the schemes for a fair comparison between their BER performances. The simulation assumptions are outlined in TABLE II in the Appendix. From Fig. 4, it is evident that under large delay spread scenario, the sequence based structure marginally outperforms DMRS based structure in terms of BER of missed ACK (i.e. combined error rate for ACK-to-NACK and ACK-to-DTX).

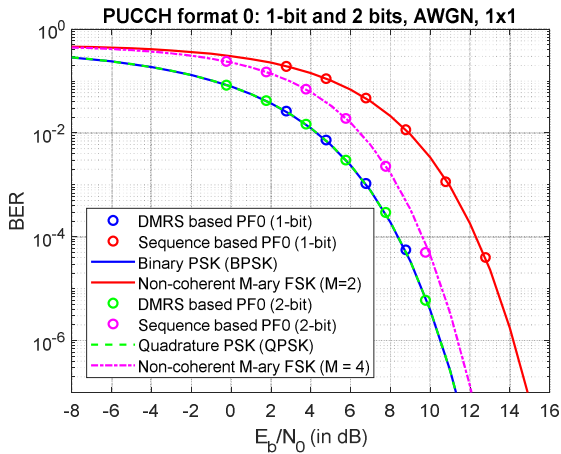

Fig. 3. BER as a function of SNR-per-bit ($E_b/N_0$)

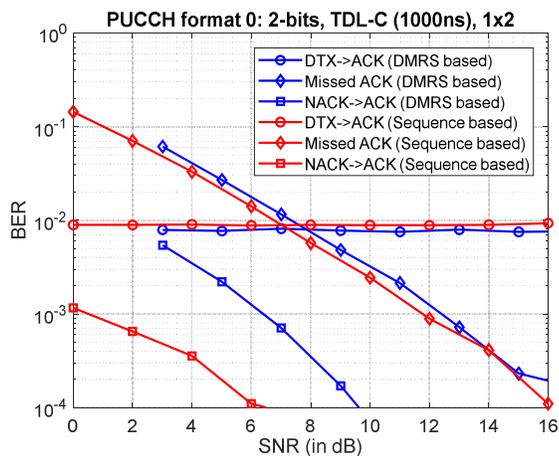

Fig. 4. Link level simulation results for PUCCH format 0

Note that, the false alarm rate (i.e. NACK-to-ACK error rate) is significantly less for sequence based structure, compared to DMRS based structure, which is an important aspect of UL control channel design.

The general observation is that, while these two structures offer comparable BER performance at small to medium delay spread scenarios, sequence based structure delivers better BER performance (i.e. lower missed ACK rate) under high delay spread scenario, a trend that continues with increase in Sub-Carrier Spacing (SCS). This is due to the fact that frequency-selectivity becomes more severe with increase in SCS, leading to further degradation in channel estimation. In the sequence based structure, orthogonal sequences that are robust under high frequency-selectivity can be employed.

Both DMRS and sequence based structures were thoroughly studied and based on the BER performance comparison between these two structures [7], sequence based design was adopted for PF0 in 3GPP Release 15 specification of NR [4][8].

*B. PUCCH Format 2*

*a) PUCCH Format 2 Structure:* PF2 is utilized to carry more than 2 UCI bits, e.g., CSI report and more than 2 HARQ-ACK feedback bits. When PUCCH format 2 conveys relatively large payload, multiple PRBs (up to 16) can be allocated for the transmission of PF2.

For PF2, after channel coding and QPSK modulation, the modulated UCI symbols are directly mapped on the allocated resource. Further, UE (User Equipment) specific DMRS is embedded for PF2 to allow gNB to perform coherent detection. This FDM based multiplexing of DMRS and UCI symbols indicates that PF2 is based on Cyclic Prefix Orthogonal Frequency Division Multiplexing (CP-OFDM) waveform. Note that DMRS is generated based on a pseudo-random sequence, where the initialization seed is defined as a function of symbol and slot indices and a configurable ID. This can facilitate to randomize the inter-cell interference on a symbol level basis for short PUCCH transmission.

*b) Simulation of DMRS Pattern and Overhead:* In this subsection, we provide link-level simulation results to analyze various DMRS overheads and patterns for PF2, which are illustrated in Fig. 5.

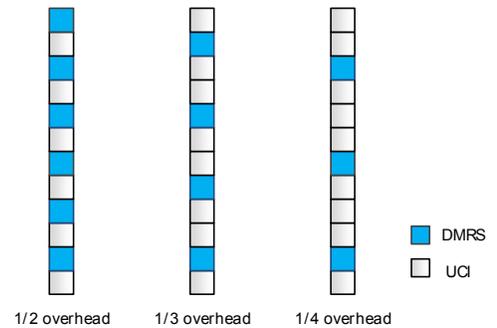

Fig. 5. DMRS patterns for PUCCH format 2

In the simulation, it is assumed that 20 bits of payload is carried by PF2, which occupies 4 PRBs. In addition, rate matching is applied after encoding to fill in all available REs allocated for UCI symbols. Other simulation parameters are listed in TABLE II in the Appendix.

Figure 6 illustrates link level simulation results for PF2. From the figure, it can be observed that DMRS with 1/3 or 1/4 overhead can achieve ~1.0 dB better link level performance compared to DMRS with 1/2 overhead. Further, DMRS with 1/3 overhead can slightly outperform DMRS with 1/4 overhead for 20 bits of payload in case of relatively large delay spread.

Based on the link level simulation results, the DMRS with evenly distributed pattern and 1/3 overhead was adopted for PF2 in NR [8].

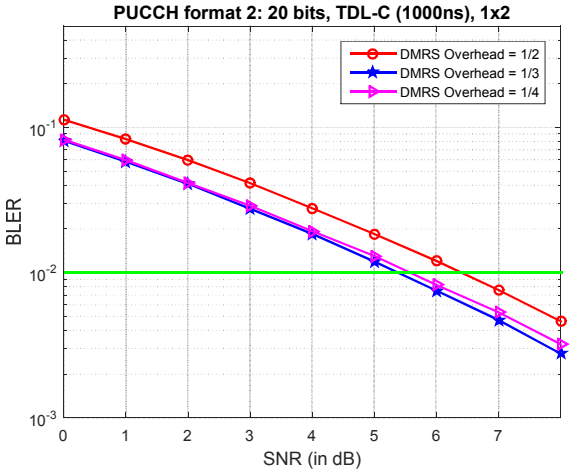

Fig. 6. Link level simulation results for PUCCH format 2

### III. LONG DURATION PUCCH

Long duration PUCCH can span 4~14 symbols within a slot and may carry either small (1~2 bits) or large (more than 2 bits) payload size. Long PUCCH carrying small payload size is referred to *PUCCH format 1* (PF 1), while the one carrying large payload size is called *PUCCH format 3* (PF 3) or *PUCCH format 4* (PF 4), depending on UE multiplexing capacity.

While PF3 has no multiplexing capacity, PF4 can multiplex up to 4 UEs on the same time-frequency resource (i.e. on the same PRB) by employing pre- Discrete Fourier Transform (DFT) Orthogonal Cover Code (OCC) [8]. In general, in long PUCCH formats, DMRS and UCI symbols are time-division multiplexed and based on DFT spread OFDM (DFT-s-OFDM) waveform. Long duration PUCCH formats are introduced in NR (in addition to short duration PUCCH) to provide coverage enhancement, which is essential to serve power limited cell areas, especially, cell edge UEs.

The detailed structures and design approaches of long duration NR PUCCH formats, i.e. PF 1 and PF 3 are described in the following subsections. PF4 has the similar time-frequency structure as PF 3 over 1 PRB.

### A. PUCCH Format 1

NR (PF 1 spans over 4~14 symbols in a slot and carries typically 1~2 UCI bit(s), which may be either HARQ-ACK bit(s) or SR bit or both. In this format, DMRS symbols are formed by low Peak to Average Power ratio (PAPR), Computer Generated Sequence (CGS) with cyclic shift in frequency domain and Orthogonal Cover Code (OCC) [9] in time domain. UCI symbols are either BPSK (1 bit) or QPSK (2 bits) modulated and multiplied with low-PAPR computer generated sequence. Additionally, OCC [9] is applied in time domain. Two different approaches can be considered for the design of PF 1, e.g. a) extension method and b) puncturing method, as illustrated in Fig. 7.

*a) Extension Method:* In this method, any length-$n$ PF1 ($4 \leq n \leq 14$) consists of $\lfloor n/2 \rfloor$ UCI symbols and $n - \lfloor n/2 \rfloor$ DMRS symbols. DMRS and UCI symbols are interleaved in TDM manner such that DMRS symbols occupy the even indexed symbol locations within length-$n$ (the starting symbol being of index 0). One example of length-7 PF1 using extension method is shown in Fig. 7a. In this option, any length-$n$ PF1 ($n > 4$) can be conceived as an extension of length-4 PF1 (the shortest length of PF1).

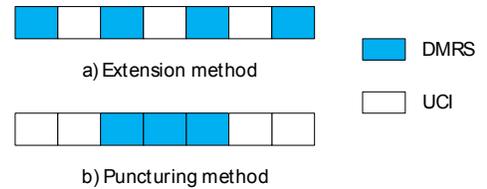

Fig. 7. Candidate structures for PUCCH format 1

*b) Puncturing Method:* In this method, DMRS symbols are bundled together and punctured into the middle of the length-$n$ PF1, while UCI symbols are placed at the beginning and end of the PUCCH length. This is similar to the structure of LTE PUCCH format 1/1a/1b [9]. An example of length-7 PF1 using puncturing method is shown in Fig. 7b. Note that, in both extension and puncturing methods, DMRS overhead is kept nearly 1/2.

*c) Simulation of Extension and Puncturing Methods:* In this subsection, we provide link level simulation results for PF1. In Fig. 8, we have plotted BER performance of three different lengths of PF1, each being constructed using extension method and puncturing method. In the simulation, it is assumed that 2-bit UCI payload is carried by PF1 with moderately high UE speed (120 km/h) and medium delay spread (100 ns).

It is observed in Fig. 8 that with increase in length of PUCCH (here, from 5 symbols to 7 symbols), the performance of extension method is consistently improving (i.e. extension method is offering lower BER) compared to puncturing method. The performance difference is more pronounced at higher UE speed. This is primarily due to the fact that in high mobility scenario, the channel across DFT-s-OFDM symbols vary more rapidly and a comb-like interleaving of DMRS and UCI in TDM manner would offer more robust channel estimation and hence better BER performance compared to the puncturing method. In puncturing method, on the other hand, the channel estimates for UCI symbols are extrapolations of DMRS channel estimates, which might not be accurate under the fast fading scenario, as is the case with high UE speed.

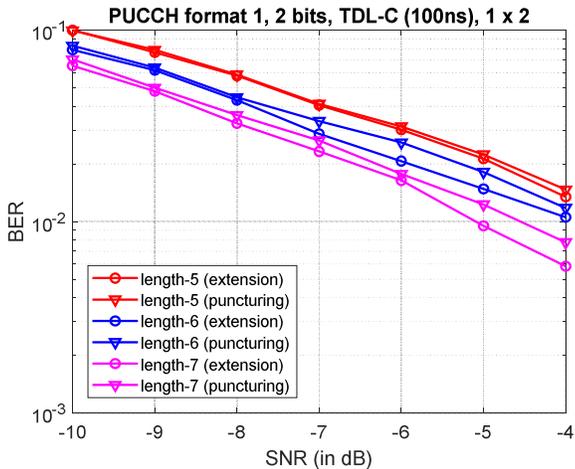
Fig. 8. Link level simulation results for PUCCH format 1

Based on the link level simulation results [10], extension method is adopted for NR PF1, where DMRS and UCI are interleaved in TDM manner and DMRS symbols occupy the even index positions across the length-$n$ of PUCCH, where the first symbol of PUCCH is indexed 0.

### B. PUCCH Format 3

NR PF3 is utilized to carry more than 2 UCI bits, e.g., CSI report and more than 2 HARQ-ACK feedback bits. When PF3 conveys relatively large payload, multiple PRBs (up to 16) can be allocated for the transmission of PF3, as indicated in TABLE I. Since UCI payload size is relatively large, the DMRS overhead for the transmission of PF3 is less than 1/2, similar to PF2.

While DMRS symbols are formed by a low PAPR, computer generated sequence in frequency domain, UCI bits are encoded, scrambled, QPSK modulated and DFT-pre-coded to form UCI symbols. In addition, $\pi/2$ BPSK is also supported as a UE specific configurable modulation scheme for UCI.

*a) PUCCH Format 3 Structure:* Similar to PF1, PF3 may span over 4~14 symbols in a slot and for length-4 PF3, 1 DMRS symbol is used to keep the DMRS overhead less than 1/2. For length-$n$ ($14 \geq n > 4$), there exists multiple design options based on number of DMRS symbols and their locations that would lead to less than 1/2 DMRS overhead.

For example, length-5 PF3 may have 1 or 2 DMRS symbols with less than 1/2 DMRS overhead, as shown in Figs. 9a and 9b. Similarly, length-10 PF3 can have any number of DMRS symbols less than 5 (e.g. 2 or 4 DMRS symbols) to maintain less than 1/2 DMRS overhead, as depicted in Figs. 9c and 9d.

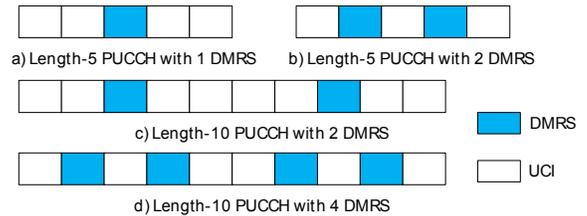
Fig. 9. Candidate structures for length-5 and length-10 PUCCH format

*b) Simulation of PUCCH Format 3:* In this subsection, we provide link level simulation results for PF3. In Figs. 10 and 11, we have plotted BER curves for length-5 and length-10 PF3 with 1 and 2 DMRS symbols, and 2 and 4 DMRS symbols respectively, with the TDM structure shown in Fig. 9.

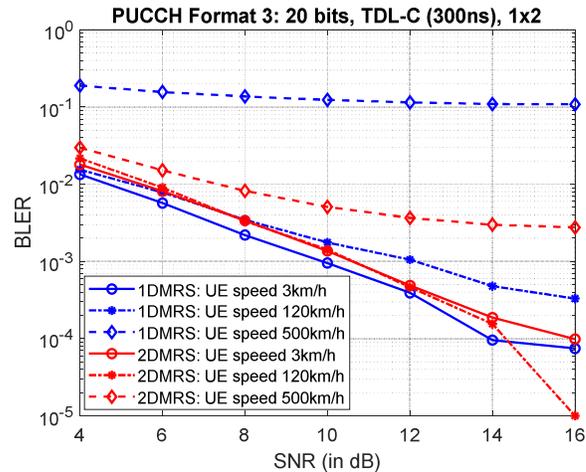
Fig. 10. Link level simulation results for length-5 PUCCH format 3

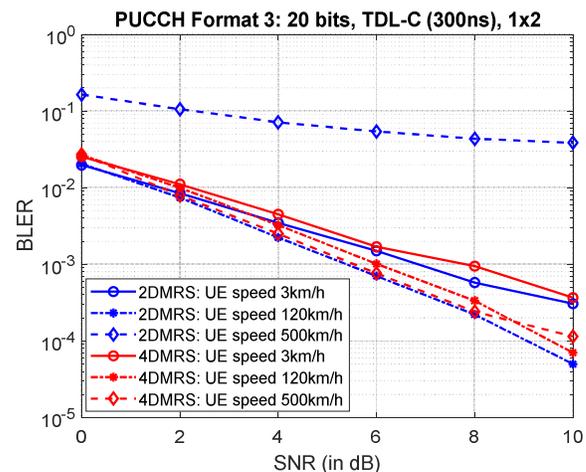
Fig. 11. Link level simulation for length-10 PUCCH format 3

In the simulation, it is assumed that 20 bits of payload is carried by length-5 and length-10 PF3. From Fig. 10, it is observed that for low UE mobility (3 km/h), 1 DMRS symbol may offer ~1dB performance gain compared to 2 DMRS symbols. With increase in UE mobility, however, 2 DMRS symbols begin to outperform 1 DMRS symbol, owing to its better channel estimation capability under high frequency selectivity scenario. Under extreme mobility situation (more than 500 km/h), 2 DMRS symbols become a necessity for length-5 PF3, since 1 DMRS symbol fails to achieve the desired BER rate of $10^{-2}$ and less.

For length-10 PF3, Fig. 11 shows that 2 DMRS symbols offer comparable BER as 4 DMRS symbols under low (3 km/h) and moderate (120 km/h) UE mobility scenarios. But under extremely high UE mobility (beyond 500 km/h), 4 DMRS symbols become a necessity to provide desired BER performance of less than $10^{-2}$.

Based on the link level simulation results [11], 2 DMRS symbols are employed for length-$n$ PF3, where $n > 4$ for NR. Moreover, for length-$n$ larger than or equal to 10 (i.e. $10 \leq n \leq 14$), 2 additional DMRS symbols (i.e. 4 DMRS symbols in total) can be configured by a signaling based on Radio Resource Control (RRC) protocol to support high UE mobility scenarios.

## IV. CONCLUSION

In this paper, we present the design principles of various PUCCH formats adopted in 3GPP NR and the underlying physical structures. With link level evaluations, we explain the technical background related to development and adoption of each of the PUCCH formats in 3GPP NR. Detailed structure of the NR PUCCH formats from a single UE perspective is analyzed. Other aspects of NR PUCCH, e.g. multiplexing of multiple UCIs on a single PUCCH resource, UE multiplexing capacity of different PUCCH formats, resource allocation of PUCCH, frequency hopping, multi-slot PUCCH transmission, transmit diversity schemes for PUCCH etc., will be addressed in our future work.

## APPENDIX

TABLE II. SIMULATION PARAMETERS

| Parameters | Configuration |
| --- | --- |
| Carrier frequency | 4 GHz |
| System bandwidth | 20 MHz |
| Subcarrier spacing | 15 KHz |
| Number of transmit antennas | 1 |
| Number of receive antennas | 2 |
| Channel model | TDL-C |
| Delay spread | 100ns, 300ns, 1000ns |
| UE velocity | 3km/h, 120km/h, 500km/h |
| Channel estimation | MMSE |
| Payload size (without CRC) | 2, 20 bits |
| Number of PRBs | 1 (PF0/1/3), 4 (PF2) |
| Number of UEs | 1 |